\documentclass[sigconf]{acmart}
\renewcommand\footnotetextcopyrightpermission[1]{} 
\usepackage[utf8]{inputenc}
\usepackage{subfig}
\usepackage{float}
\usepackage{stfloats}
\usepackage{url}
\usepackage{multirow}
\usepackage{tablefootnote}
\usepackage{multicol}
\usepackage{hyperref}
\usepackage{graphicx}
\usepackage{booktabs}

\usepackage[T1]{fontenc}

\usepackage{boldline}
\usepackage[toc]{appendix}
\usepackage{geometry}

\usepackage[T1]{fontenc}

\usepackage{color}

\definecolor{pblue}{rgb}{0.13,0.13,1}
\definecolor{pgreen}{rgb}{0,0.5,0}
\definecolor{pred}{rgb}{0.9,0,0}
\definecolor{pgrey}{rgb}{0.46,0.45,0.48}

\usepackage{listings}
\AtBeginDocument{%
  \providecommand\BibTeX{{%
    \normalfont B\kern-0.5em{\scshape i\kern-0.25em b}\kern-0.8em\TeX}}}

\acmConference[Analytics and Machine Learning for Software Engineering]{}{2019}{TU Delft}
\acmBooktitle{Final Paper}
\acmPrice{0}
\acmISBN{ }

\begin{document}

\title[]{Using Distributed Representation of Code for Bug Detection 
}

\author{Jón Arnar Briem$^*$}
\affiliation{
  4937864 \\
  TU Delft
}
\email{j.a.briem@student.tudeft.nl}

\author{Jordi Smit$^*$}
\affiliation{
  4457714 \\
  TU Delft
}
\email{j.smit-6@student.tudeft.nl}

\author{Hendrig Sellik$^*$}
\affiliation{
  4894502 \\
  TU Delft
}
\email{h.sellik@student.tudeft.nl}

\author{Pavel Rapoport$^*$}
\affiliation{
  4729889 \\
  TU Delft
}
\email{p.rapoport@student.tudeft.nl}

\begin{abstract}
    Recent advances in neural modeling for bug detection have been very promising. More specifically, using snippets of code to create continuous vectors or \textit{embeddings} has been shown to be very good at method name prediction and claimed to be efficient at other tasks, such as bug detection. However, to this end, the method has not been empirically tested for the latter.

    In this work, we use the Code2Vec model of Alon et al. to evaluate it for detecting off-by-one errors in Java source code. We define bug detection as a binary classification problem and train our model on a large Java file corpus containing likely correct code. In order to properly classify incorrect code, the model needs to be trained on false examples as well. To achieve this, we create likely incorrect code by making simple mutations to the original corpus.
    
    Our quantitative and qualitative evaluations show that an attention-based model that uses a structural representation of code can be indeed successfully used for other tasks than method naming. 
\end{abstract}


\maketitle
\def\thefootnote{*}\footnotetext{The authors contributed equally to this work}\def\thefootnote{\arabic{footnote}}

\section{Introduction}
The codebases of software products have increased yearly, now spanning to millions making it hard to grasp the knowledge of the projects \cite{mcmillan2013portfolio}. All this
code needs to be rapidly developed and also maintained. As the amount of code is tremendous, it is an easy opportunity for bugs to slip in.

There are multiple tools that can help with this issue. Most of the time, working professionals rely on static code analyzers such as the ones found in IntelliJ IDEA\footnote{\url{https://www.jetbrains.com/idea/}}. However, these contain a lot of false positives or miss the bugs (such as most off-by-one errors) which make the developers ignore the results of the tools \cite{goseva2015capability}. Recently, a lot of Artificial Intelligence solutions have emerged which help with the issue \cite{alon2019code2vec, pradel2018deepbugs, allamanis2017learning, vasic2019neural}. Although they are not a panacea, they provide enhanced aid to developers in different steps of developing processes, highlighting the potentially faulty code before it gets into production.

One particular solution by Alon et al., the Code2Vec model \cite{alon2019code2vec} (see Section \ref{sec:relevant_literature} for description), delivers state-of-the-art performance on method naming. The authors do not test the model on other tasks, however, they theorize that it should also yield good performance. 

\begin{figure*} 
  \includegraphics[width=\textwidth,height=8cm]{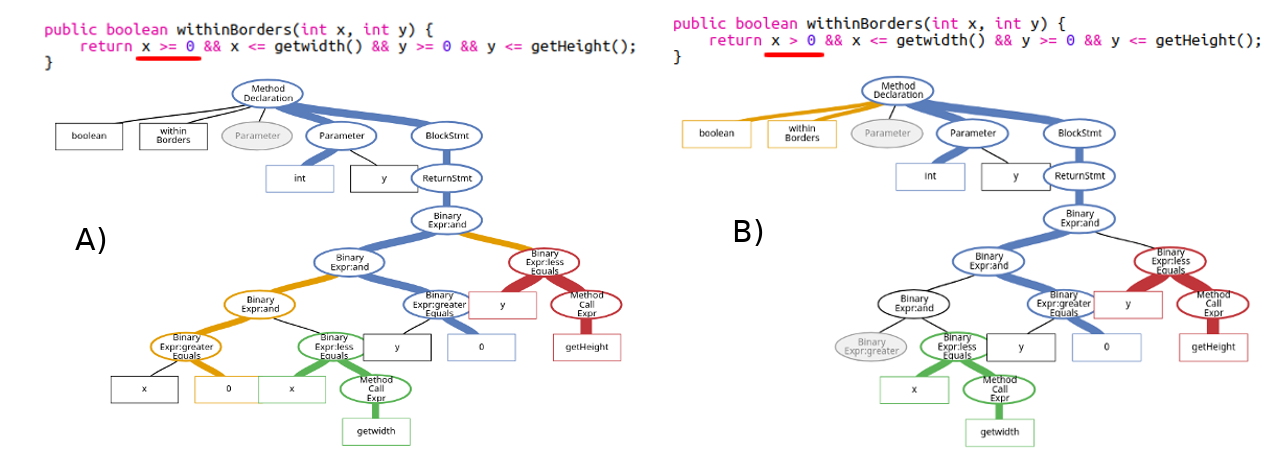}
  \caption[]{Example change of AST weights after introducing bug using Code2Vec web interface\protect\footnotemark} (Option \textit{A} being correct and \textit{B} incorrect code)
  \label{fig:AST_exclamation}
\end{figure*}

In this work, we use the Code2Vec deep learning model and replace the layer for method naming with a binary classification layer. The aim is to repurpose the model for detecting off-by-one logic errors found in boundary conditions (see Section \ref{sec:data}). Our intuition is that the change in the Abstract Syntax Tree (AST) weights upon introducing a bug as seen in Figure \ref{fig:AST_exclamation} is learnable by our model. Hence the system should be able to learn to identify off-by-one bugs.

The main contributions of this paper are.
\begin{itemize}
    \item Replicating work done by authors of Alon et al. \cite{alon2019code2vec}
    \item Quantitative and qualitative evaluation of a Code2Vec's performance on a task other than method naming.
\end{itemize}

The paper is divided into the following sections. In Section \ref{sec:relevant_literature} the relevant literature used to create this paper is discussed, in Section \ref{sec:data} the data origins and preprocessing is explained, in Section \ref{sec:model} the architecture and training of the model are discussed. Finally, in Section \ref{sec:evaluation}, the model is evaluated which is followed by some reflections regarding our work in Section \ref{sec:discussion} and a conclusion in Section \ref{sec:conclusion}.

\footnotetext{\url{https://code2vec.org/}}

\section{Relevant Literature} \label{sec:relevant_literature}
\textit{Code2Vec} by Alon et al. \cite{alon2019code2vec} is a state-of-the-art deep neural network model used to create fixed-length vector representations (\textit{embeddings}) from code. The embeddings of similar code snippets encode the semantic meaning of the code, meaning that similar code has similar embeddings. 

The model relies on the paths of the Abstract Syntax Tree (AST) of the code. While the idea of using AST paths to generate code embeddings is also used by other authors such as Hu et al. \cite{hu2018deep}, Code2Vec is superior due to the novel way of using AST paths. Instead of linearizing paths as \cite{hu2018deep}, Code2Vec authors use a path-attention network to identify the most important paths and learn the representation of each path while simultaneously learning how to aggregate a set of them.

The authors make use of embedding similarity of similar code to predict method names. The model is trained on a dataset of 12 million Java methods and compared to other competitive approaches. It significantly outperforms them by having approximately 100 times faster prediction rate at the same time having a better F1 score of 59.5 at method naming. While they only tested their model for method naming, the authors believe that there are a plethora of programming language processing tasks the model can be used for.

\textit{DeepBugs} by Pradel et al. \cite{pradel2018deepbugs} uses a deep learning model to identify bugs related to swapped function arguments, wrong binary operators and wrong operands in binary operation. The model creates an embedding from the code, but instead of using ASTs like Code2Vec, the embedding is created from specific parts of the code. For example, embeddings for identifying swapped function arguments are created from the name of the function, names and types of the first and second arguments to the function with their parameter names, and name of the base object that calls the function.

They use a Javascript dataset containing likely correct code. In order to generate negative examples for training, they use simple transformations to create likely incorrect code. The authors formulate bug detection as a binary classification problem and evaluate the embeddings quantitatively and qualitatively. The approach yields an effective accuracy between 89.06\% and 94.70\% depending on the problem at hand.

\section{Data} \label{sec:data}
In this section, we describe the data used to train our model. In Section \ref{sec:obo_errors} we first describe the type of bug we aim to detect, in Section \ref{sec:datasets} we describe the data we used. In Section \ref{sec:mutations} we describe how we mutated the data to generate positive code (containing bugs) and finally we describe how we represent source code as a vector using the approach defined by Alon et al. \cite{alon2019code2vec} in Section \ref{sec:SourceCodeRepresentation}.

\subsection{Off-by-one errors}\label{sec:obo_errors}
Off-by-one errors (in Java terms '<' vs '<=' and '>' vs '>=') are generally considered to be among the most common bugs in software. They are particularly difficult to find in source code because there exist no tools which can accurately locate them and the result is not always obviously wrong, it is merely off-by-one. In most cases, it will lead to an "Out of bounds" situation which will result in an application crash. However, it might not crash an application and might lead to arbitrary code execution or memory corruptions, potentially exploitable by adversaries \cite{dowd2006art}. 

Manually inspecting code for off-by-one errors is very time-consuming since determining which binary operator is actually the correct one is usually heavily context-dependent. This is also why we chose to base our approach on the work done by Code2Vec \cite{alon2019code2vec}, this allows the model to discover code context heuristics which static code analyzers are not able to capture. 

\subsection{Datasets}\label{sec:datasets}

To train and validate our project we use the java-large dataset collected by Alon et al. \cite{code2seq}. It consists of 1000 top-starred projects from GitHub and contains about 4 million examples. Furthermore, since the Code2Vec model was already trained on this data we reduce the transfer learning needed to adapt the weights between the two models. 

The distributions of '<', '<=', '>' or '<=' (hereafter referred to as \textit{comparators}) and the types of statements containing those comparators in the original dataset can be seen in tables \ref{tab:ComparatorDistribution} and \ref{tab:StatementDistribution}. As can be seen, the distributions of both comparators and statement types is far from uniform with < accounting for over half of the comparators and \textit{if} statements containing a similar share of all comparators. We considered trying to balance the dataset and see what effect that would have on the accuracy of the model but were unable to do so due to time constraints.

\begin{table}[]
\centering
\begin{tabular}{l|l|l}
\toprule
Comparator    & Count   & Percentage \\ 
\midrule
greater       & 755,078  & 27.43\%    \\ 
greaterEquals & 35,9455  & 13.06\%    \\ 
less          & 1,389,789 & 50.48\%    \\ 
lessEquals    & 248,848  & 9.04\%     \\ 
\bottomrule
\end{tabular}
\caption{Distribution of comparators found in the java-large dataset collected by Alon et al.\cite{code2seq}}
\end{table}
\label{tab:ComparatorDistribution}

\begin{table}[]
\centering
\begin{tabular}{l|l|l}
\toprule
Statement Type    & Count   & Percentage \\ 
\midrule
If             & 1,399,436 & 50.83\%    \\
For            & 9,219,48  & 33.49\%    \\
While          & 104,412  & 3.79\%     \\
Ternary        & 100,567  & 3.65\%     \\
Method         & 70,268   & 2.55\%     \\
Other          & 68,735   & 2.50\%     \\
\bottomrule
\end{tabular}
\caption{Distribution of the type of statements containing comparators found in the java-large dataset collected by Alon et al.\cite{code2seq}}
\end{table}
\label{tab:StatementDistribution}

For our training data we need both positive (containing bugs) code and negative code (bug-free), to do this we take the aforementioned dataset and create bugs by performing specially designed mutations. First, we parse the code using Java Parser \footnote{https://javaparser.org/} to generate an AST based on the source code. We then search the AST for methods that might contain the bug in question, generate a negative example by performing the mutation, which will be further explained in the next section. Then we take both the negative and positive examples and send them through the preprocessing pipeline used in \cite{alon2019code2vec} before passing them into our model. This process is described in detail in the next sections.

\subsection{Mutations}\label{sec:mutations}

To generate our mutations we consider each method and search for any occurrences of comparators. If the method contains a comparator it is extracted from the code for further manipulation, any methods not containing a comparator are ignored. Once we have extracted all methods of interest from the code, the method is parsed for different \textit{context types} that comparators occur in. This is done so that we can account for different sub-types of off-by-one errors and also evaluate if our model is better at detecting bugs in some contexts than others. 

\begin{lstlisting}[caption={An example of a method before mutation. The context type of this comparator is FORless},captionpos=b]
public void setContents(List<Content> contentsBefore, List<Content> contentsAfter) {
    for (int i = 0; i < contentsAfter.size(); i++) {
        Content content = contentsAfter.get(i);
        if (content instanceof PathContent) {
            paths.add((PathContent) content);
        }
    }
}
\end{lstlisting}

We define a context type as the combination of the operator (less, lessEquals, greater or greaterEquals) and the type of statement it occurs in (For loop, If statement, While loop, etc), more specifically this is the class of the operator's parent node in the AST. For a full list of all context types refer to Appendix \ref{appendix:quantitative_evaluation}. For every method, a random comparator is selected, mutated and turned into its respective off-by-one comparator (for example, '<' will be mutated into '<='). Finally both the original and the mutated methods are added to the dataset.

\begin{lstlisting}[caption={An example of a method after mutation. The context type of this comparator is FORlessEqueals},captionpos=b]
public void setContents(List<Content> contentsBefore, List<Content> contentsAfter) {
    for (int i = 0; i <= contentsAfter.size(); i++) {
        Content content = contentsAfter.get(i);
        if (content instanceof PathContent) {
            paths.add((PathContent) content);
        }
    }
}
\end{lstlisting}

\subsection{Source code representation} \label{sec:SourceCodeRepresentation}
After a method has been extracted, it is passed through the same preprocessing pipeline as was used in the original Code2Vec paper \cite{alon2019code2vec}. The source code of each method is again turned into an AST using the modified JavaExtractor from \cite{alon2019code2vec}. We then select at most $200$ paths between $2$ unique terminals in the AST of the method. We encode these terminals into integer tokens using the dictionary used by Code2Vec \cite{alon2019code2vec} and hash the string representation of the paths with Java hashcode method\footnote{\url{https://docs.oracle.com/javase/8/docs/api/java/lang/String.html\#hashCode--}}. This means that each method in Java code is turned into a set of at most $200$ integer tuples of the format $(terminal_i, path, terminal_j)$ whereby $i \neq j$ and $path$ is an existing path in the AST between source $terminal_i$ and target $terminal_j$. 

In this paper, we want to apply transfer learning from the original Code2Vec model \cite{alon2019code2vec} to our problem. Allowing us to reuse the pre-trained weights from the original Code2Vec model \cite{alon2019code2vec}. However, we can only do this if our model preprocesses and encodes the source code in the same way as described in the original Code2Vec paper \cite{alon2019code2vec}. That is why our preprocessing code is based on the Java Extractor Jar and dictionaries as published on the repository \footnote{https://github.com/tech-srl/code2vec} of Code2Vec and tailored to our needs.

\section{Model} \label{sec:model}
In this section, we describe our model in detail. Section \ref{sec:NeuralNetworkArchitecture} describes the architecture of the neural network. Section \ref{sec:Training} describes how the network was trained. Finally, we cover the precision of our model in Section \ref{sec:modelPrecision}.

\subsection{Neural Network Architecture}\label{sec:NeuralNetworkArchitecture}
\begin{figure*}[b] 
  \includegraphics[width=\textwidth,height=7cm]{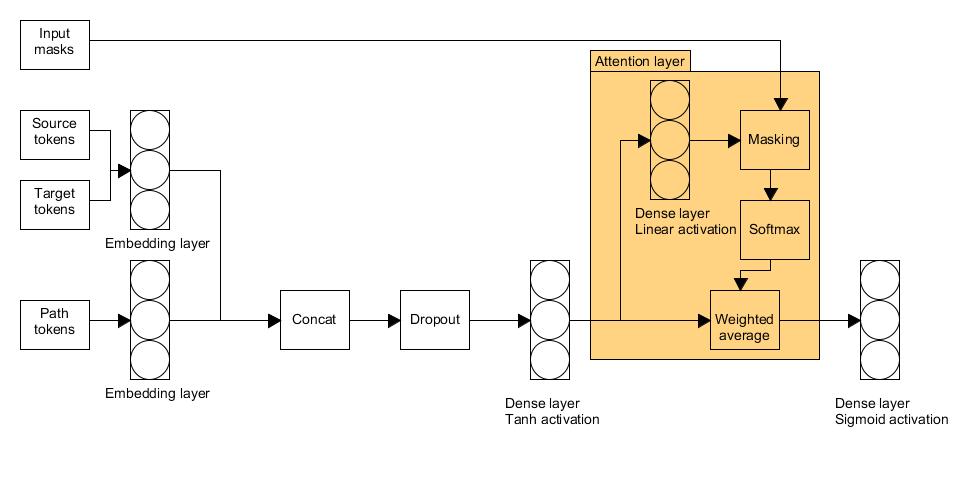}
  \caption{Neural network architecture}
  \label{fig:networkArchitecture}
\end{figure*}

The model is an attention-based model based on Code2Vec \cite{alon2019code2vec} whereby the overwhelming majority of the weights are in the embedding layer of the network. The architecture of the model has been depicted in Figure \ref{fig:networkArchitecture}. The model takes a set of at most $200$ integer token tuples of the format $(terminal_i, path, terminal_j)$ as an input. It embeds these inputs into a vector with $128$ parameters, whereby the terminal tokens and the path tokens each have their own embedding layer. These embeddings are concatenated into a single vector and passed through a dropout layer. These $200$ vectors are then combined  into a single context vector using an attention mechanism. The context vector will be used to make the final prediction. 

Our model's architecture is similar to the architecture of the original Code2Vec model \cite{alon2019code2vec}. The only difference is in the final layer where our model uses Sigmoid activation with a single output unit. This is because we have a binary classification problem instead of a multi-classification problem. We chose to use the same architecture as Code2Vec to allow the use of transfer learning using the pre-trained weights of Code2Vec \cite{alon2019code2vec}. This also allows us to verify the claim made by the Code2Vec authors that there are a plethora of
programming language processing tasks that their model can be used for \cite{alon2019code2vec}.

\subsection{Training}\label{sec:Training}
For the training and validation process, we preprocessed the raw Java training and validation dataset collected by Alon et al. \cite{code2seq}. The generation process resulted in a training set of $1,512,785$ data points (mutated or correct Java methods) and a validation set of $28,086$ data points that were used to train and test our model respectively. We used large over the medium dataset \cite{code2seq} because it resulted in slightly higher overall scores (Table \ref{tab:evaluationDataset}). 
For this training process, we used binary cross-entropy as our loss function and Adam \cite{kingma2014adam} as our optimization algorithm. The model was also trained using early stopping on the validation set. The training process was halted after the accuracy on the validation set did not increase for $2$ Epochs and the weights with the lowest validation loss were kept.

The authors of Code2Vec have speculated that their pre-trained weights could be used for transfer learning \cite{alon2019code2vec}. That is why we experimented with applying transfer learning in two ways. Firstly, we attempted Feature Extraction whereby the pre-trained weights of the Code2Vec model were frozen and only the final layer was replaced and made trainable. Secondly, we tried Fine-Tuning with pre-trained weights of the Code2Vec model as the initial value of our model and allowed the model to update all the values as it saw fit, expect the embeddings weights. Finally, we also trained a model with randomly initialized weights as a baseline. The resulting accuracies are displayed in table \ref{tab:evaluationArchitectures}.

\begin{table}[]
\begin{tabular}{@{}lllll@{}}
\toprule
Model     & \multicolumn{1}{c}{\begin{tabular}[c]{@{}c@{}}Feature\\ Extraction\end{tabular}} & \multicolumn{1}{c}{Fine-Tuning} & \multicolumn{1}{c}{\begin{tabular}[c]{@{}c@{}}Randomly\\ Initialized\end{tabular}} & \multicolumn{1}{c}{\begin{tabular}[c]{@{}c@{}}Transformer \\ model\end{tabular}} \\ \midrule
Accuracy  & 0.732                                                                            & 0.788                           & \textbf{0.790}                                                                     & 0.717                                                                            \\
F1        & 0.709                                                                            & \textbf{0.781}                  & 0.777                                                                              & 0.674                                                                            \\
Precision & 0.776                                                                            & 0.809                           & \textbf{0.831}                                                                     & 0.794                                                                            \\
Recall    & 0.653                                                                            & \textbf{0.756}                  & 0.730                                                                              & 0.585                                                                            \\ \bottomrule
\end{tabular}
\caption{Evaluation comparison of different architectures on an unseen test of $88,228$ examples based on the mutated test set from Alon et al. \cite{code2seq}}
\label{tab:evaluationArchitectures}
\end{table}

\begin{table}[]
\centering
\begin{tabular}{@{}lll@{}}
\toprule
Model     & \begin{tabular}[c]{@{}l@{}}Fine-Tuning\\ medium dataset\end{tabular} & \multicolumn{1}{c}{\begin{tabular}[c]{@{}c@{}}Fine-Tuning\\ large dataset\end{tabular}} \\ \midrule
Accuracy  & 0.758                                                                & \textbf{0.788}                                                                          \\
F1        & 0.757                                                                & \textbf{0.781}                                                                          \\
Precision & 0.762                                                                & \textbf{0.809}                                                                          \\
Recall    & 0.751                                                                & \textbf{0.756}                                                                          \\ \bottomrule
\end{tabular}
\caption{Evaluation comparison of the medium data set  ($482,138$ training examples) and the large data set ($1,512,785$ training examples)  \cite{code2seq} .}
\label{tab:evaluationDataset}
\end{table}

\subsection{Alternative Architecture}\label{sec:AlternativeNeuralNetworkArchitecture}
Besides the original Code2Vec architecture, we also tried a more complicated architecture based on BERT \cite{bertdevlin2018bert}. The idea behind this architecture was that the power of the Code2Vec \cite{alon2019code2vec} architecture lays in the attention mechanism. To verify this idea we tried a transformer model based on BERT. This model encodes the source cod,e in the same manner as the Code2Vec model. It then embeds these input tokens using the frozen pre-trained embedding layers from Code2Vec. Finally, it passes these embedding through multiple BERT layers before making the final prediction. 
However as can be seen in Table \ref{tab:evaluationArchitectures}, this architecture achieved a much lower overall score than the original Code2Vec architecture. This is most likely due to the observed Java code AST being limited to method scope as mentioned in Section \ref{sec:data}. Since our preprocessing is highly depended on the Code2Vec preprocessing pipeline and the overall scores for this model were lower, we did not pursue this architecture any further.

\subsection{Predicting Bug Location}\label{sec:modelPrecision}
Our model looks for bugs with a method-level precision, that is, it predicts whether a method contains an off-by-one bug or not. It is not able to detect where exactly inside that method the bug is. This is due to our effort to capture the context of the whole method. Looking at a smaller section of code would yield more precisely located bugs, but possibly at the cost of context which would reduce the accuracy of the model.

\section{Evaluation} \label{sec:evaluation}
In this section, we present the results for our quantitative analysis to better understand our model and select the best performing architecture. We also perform a qualitative evaluation with the best performing architecture and compare it to static analyzers.
\subsection{Quantitative Evaluation} \label{sec:quantitative_analysis}
Different model architectures were compared with regards to simple metrics such as accuracy, F1 score, precision and recall. Out of the 4 tested architectures, the fine-tuning Code2Vec architecture was selected for further evaluation based on the metrics (see Table \ref{tab:evaluationArchitectures}). This architecture is also used in the qualitative evaluation.

To better understand the model and to analyze the context types in which it performs the best, we broke down the performance of the model by context type (see Table \ref{tab:quantitative_evaluation_bug_types} in Appendix \ref{appendix:quantitative_evaluation}). The test dataset is previously unseen code from the same data-gathering project and mutation method as used to train the model (see Section \ref{sec:data}), hence the distribution of different bug opportunities remains the same.

From Table \ref{tab:quantitative_evaluation_bug_types}, it is evident that the precision is correlated with the total amount of data points available for each context type and the types which have the highest number of occurrences also tend to produce a higher F1 score. For example, our model achieves an F1 score of 0.87 when detecting bugs in \textit{for loops}, which are well represented in our dataset. However, our model only achieves an F1 score of 0.52 detecting bugs when \textit{assigning} a boolean value to a variable with a logical condition. A case that is severely underrepresented in the training data.

It is notable that the model can also perform well with off-by-one errors in moderately underrepresented classes such as \textit{return statements} (F1 score of 0.73)  and \textit{while loops} (F1 score of 0.71). This might mean that there was just enough data or the problems were similar enough for the off-by-one errors in \textit{if statements} and \textit{for loops} for the model to generalize. The most underrepresented classes like \textit{assigning value} to a variable are noisy and the model was not able to generalize towards those classes.

In order to get more detailed insight, the 12 cases were further divided into 48 (see Table \ref{tab:quantitative_evaluation_all_bugs} in Appendix \ref{appendix:quantitative_evaluation}). A total of 104 958 data points (Java methods) were created for testing. Each data point contains a method with a comparator and a possibility for an off-by-one error. The names in the first column of Table \ref{tab:quantitative_evaluation_all_bugs} indicate the class of the off-by-one error (for example \textit{FOR} stands for \textit{for loop}) and the comparator indicates which comparator was passed to the model (for example \textit{less} stands for <). The comparator can be from the original code (hence likely correct), but it may also be introduced by a mutation (hence likely incorrect). Hence \textit{FORless} is a method containing a \textit{for loop} with a \textit{<} operator which could have been mutated or not. We can then compare the model output with the truth.

The data points were passed into the model which gave a prediction for the data point to be a bug (true) or not a bug (false). The detailed analysis shows that the classical \textit{FOR} loop (with < operator) scores are significantly higher than others (86-89\% accuracy). This might be due to for loops with comparators such as \textit{(int i = 0; i < number; i++) } being considered as a boilerplate in Java code.

It can also be seen that the model is biased towards predicting \textit{<=} in a for a loop as a bug and \textit{<} not as a bug. This can be explained by the balance of the training set where the majority of the for loops contain a \textit{<} operator. Hence model learns to classify our mutated code with \textit{<=} operator as faulty.

The results with \textit{if conditions} seemed fascinating from Table \ref{tab:quantitative_evaluation_all_bugs} because there is not a default structure as there is with \textit{for loops} and knowledge of the context is needed to make a prediction. Hence we trained a model specifically with mutations in \textit{if conditions} to see how it will perform on the test data.

The results of this specific model can be seen in Table \ref{tab:quantitative_if_evaluation} in Appendix \ref{appendix:quantitative_evaluation}. Interestingly, the results are not better than the model trained with all mutations and the model performs worse when judging \textit{if statements} that contain a \textit{>} operator (F1 score 0.63 vs 0.34). One possible reason for this might be that the model can generalize the relationship better with more data, independent of the contexts like \textit{if conditions} or \textit{for loops}.

\subsection{Qualitative Evaluation}
While running our model to find bugs on Apache Tomcat\footnote{\url{https://github.com/apache/tomcat}}, Apache Ant\footnote{\url{https://github.com/apache/ant}} and Apache Druid\footnote{\url{https://github.com/apache/incubator-druid}} projects, we were not able to find bugs that broke functionality, but problems that were related to code quality or false positives  (see table \ref{tab:manualEvaluationTable1} in appendix \ref{appendix:qualitative_evaluation}). This also correlates with the results of Ayewah et al. \cite{ayewah2010google} who found that most important bugs in production code are already fixed with expensive methods such as manual testing or user feedback. A much more realistic use-case for such tools is in the developing stage, where the benefit is the largest \cite{johnson2013don}. Hence we analyze code snippets concerned with off-by-one errors and highlight the situations in which the model succeeds over static analyzers and the situations it does not. 

We used 3 different static analyzers as a baseline for our evaluation. \textbf{SpotBugs} (v.4.0.0-beta1)\footnote{SpotBugs official GitHub page: \url{https://github.com/spotbugs/spotbugs}}, formerly known as FindBugs\cite{hovemeyer2004finding}, is an open-source static code analyzer for Java. It analyzes Java bytecode for occurrences of different patterns that are likely containing a bug. At the time of writing this report, SpotBugs is able to identify over 400 of such patterns\footnote{\url{https://spotbugs.readthedocs.io/en/latest/bugDescriptions.html}}, out of which 6 are relevant for this report (see table \ref{tab:patternsUsed}).

\begin{table*}[!htpb]
\begin{tabular}{@{}lll@{}}
 
\toprule
\multicolumn{1}{c}{Tool} & \multicolumn{1}{c}{\begin{tabular}[c]{@{}c@{}}Pattern\\ Code\end{tabular}} & \multicolumn{1}{c}{Meaning} \\ \midrule
\multicolumn{1}{|c|}{\textbf{PVS-Studio}} & \multicolumn{1}{l|}{V6003} & \multicolumn{1}{l|}{\begin{tabular}[c]{@{}l@{}}The analyzer has detected a potential error \\ in a construct consisting of conditional statements.\end{tabular}} \\ \midrule
\multicolumn{1}{l|}{} & \multicolumn{1}{l|}{V6025} & \multicolumn{1}{l|}{\begin{tabular}[c]{@{}l@{}}When indexing into a variable of type 'array', 'list', \\ or 'string', an 'IndexOutOfBoundsException' exception \\ may be thrown if the index value is outbound the valid range. \\ The analyzer can detect some of such errors.\end{tabular}} \\ \midrule
\multicolumn{1}{|c|}{\textbf{SpotBugs}} & \multicolumn{1}{l|}{IL\_INFINITE\_LOOP} & \multicolumn{1}{l|}{An apparent infinite loop} \\ \midrule
\multicolumn{1}{l|}{} & \multicolumn{1}{l|}{RpC\_REPEATED\_CONDITIONAL\_TEST} & \multicolumn{1}{l|}{Repeated conditional tests} \\ \cmidrule(l){2-3} 
\multicolumn{1}{l|}{} & \multicolumn{1}{l|}{RANGE\_ARRAY\_INDEX} & \multicolumn{1}{l|}{Array index is out of bounds} \\ \cmidrule(l){2-3} 
\multicolumn{1}{l|}{} & \multicolumn{1}{l|}{RANGE\_ARRAY\_OFFSET} & \multicolumn{1}{l|}{Array offset is out of bounds} \\ \cmidrule(l){2-3} 
\multicolumn{1}{l|}{} & \multicolumn{1}{l|}{RANGE\_ARRAY\_LENGTH} & \multicolumn{1}{l|}{Array length is out of bounds} \\ \cmidrule(l){2-3} 
\multicolumn{1}{l|}{} & \multicolumn{1}{l|}{RANGE\_STRING\_INDEX} & \multicolumn{1}{l|}{String index is out of bounds} \\ \bottomrule
\end{tabular}
\caption{Static analysis tools and their specific patterns used for comparison.}
\label{tab:patternsUsed}
\end{table*}

Secondly, we used \textbf{PVS-Studio} (v.7.04.34029)\footnote{PVS-Studio official home page: \url{https://www.viva64.com/en/pvs-studio/}} which is a proprietary static code analysis tool for programs written in C, C++, C\# and Java. It is aimed to be run right after the code has been modified to detect bugs in the early stage of developing process\footnote{\url{https://www.viva64.com/en/t/0046/}}. Out of 75 possible patterns for Java code analysis\footnote{PVS-Studio Java patterns: \url{https://www.viva64.com/en/w/\#GeneralAnalysisJAVA}}, 2 were suitable for our evaluation which are listed in table \ref{tab:patternsUsed}. Thirdly, we used the static analyzer integrated into \textbf{IntelliJ IDEA} Ultimate\footnote{\url{https://www.jetbrains.com/idea/}} (v. 2019.2.3).

The results show that the static analyzer of IntelliJ and SpotBugs are not able to detect off-by-one errors even if the size of the iterated array is explicitly stated. It is possible to observe this in the first example of table \ref{tab:manualEvaluationTable1},  where our model and PVS-Studio are able to detect the issue.

However, from the same table, it can be seen, that our model has learned to identify \textit{<=} operator in for loops as a bug which confirms the results of quantitative analysis in Section \ref{sec:quantitative_analysis}. This was also the case when applying the model to GitHub projects. This could be considered as a styling issue as it is considered a best practice to loop with a \textit{<} comparator.

It can be also be seen, that the false positives with \textit{<=} do not affect other bugs, such as code analyzed in Table  \ref{tab:manualEvaluationTable2} example 2. In the example, the \textit{<=} operator is replaced with \textit{<} by mistake. Our model reports it as a bug while none of the static analyzers are able to detect this mistake. This is a case where our model outperforms static analyzers. 

\section{Reflections}\label{sec:discussion}

From the work conducted, several points can be singled out as discussing topics.

\subsection{Advantages}
\begin{itemize}
    \item {\textbf{Better performance in non-trivial cases}: static analyzers that we used for evaluation can detect only very specific cases of off-by-one error. Our model allows to predict a much greater variety of off-by-one error cases.}
    
    \item {\textbf{Versatility}: The presented model can be trained to detect not only off-by-one errors but many other kinds of bugs with relatively small changes during the preprocessing stage. Only the algorithm used to create mutations should be changed in order to train our model to detect a different bug as long as a suitable training dataset exists.}
\end{itemize}

\subsection{Issues and potential improvements}
\begin{itemize}
    \item {\textbf{Using untraditional coding style leads to false positives}: using a for-loop of type \textit{for (i = start; i < end; ++i)} is very popular and a commonly used style. As a result, our model has 'learned' that any case of a for loop using <= is most likely a bug which in reality has a high chance of being a false positive. }
    
    \item{\textbf{Unbalanced dataset}
    As stated in Section \ref{sec:datasets} our initial dataset contains 4 times fewer usages of >= or <= compared to usages of > or <. This difference can lead to biased training and as a result, our model is more tending to give false-positive results in case of >=/<= usage. One of the ways to reduce this influence is the creation of balanced dataset with more equal distribution of binary operators as well as the distribution of the places of their occurrence (if-conditions, for- and while-loops, ternary expressions, etc.)}
    
    \item {\textbf{Unknown behavior on long methods}: we currently consider at most 200 context paths. This is acceptable for our dataset where most of the methods are not very long. However, if input methods will be longer this might not be enough to provide decent predictions. The severity of this issue should be checked via further experiments on an appropriate dataset. However, it should be noted that increasing the size of context paths will also increase the computing time.}
    
    \item {\textbf{Current method constraints}: The AST paths extracted are constrained to the current method only. This issue is an artifact of the Code2Vec data extraction method. This approach is valid for their purpose since they do not need to know what is happening inside of the child methods which are called from the parent method to predict the name of the latter. However, for bug detection this knowledge is crucial and omitting contents of called methods might lead to unpredictable results. This could be solved by expanding the AST with the content of some of the inner method calls. However, as with the previous case, the maximum number of AST paths should be kept in mind.}
    
    \item {\textbf{Bug Creation}: Our dataset was created by inserting bugs into code using mutations similar to the approach used by \cite{pradel2018deepbugs}. This approach allows us to have plenty of data for training but depends on the quality of the original dataset. If that dataset contains many off-by-one errors, the model will interpret those errors as correct code resulting in lower accuracy. It is also possible that off-by-one errors that happen 'naturally' when developers write code are in some way different from the errors we created with mutations. If so, our model may not be able to correctly detect those 'natural' off-by-one bugs.}
    
\end{itemize}{}

\subsection{Future work} \label{sec:future_work}
As future research, the same method could be applied to other languages. The model should benefit more with languages with dynamic typing, such as Javascript or Python. For the latter, a context path extractor\footnote{\url{https://github.com/vovak/astminer/tree/master/astminer-cli}} was recently created by Kovalenko et al. \cite{kovalenko2019pathminer}.

It might also be interesting to see if it is possible to achieve a
higher overall score by not limiting the AST paths to the method scope. This should be possible since the model with
the randomly initialized weights achieved similar overall scores to
the model with fine-tuned weights. Hence, one is free to use altered encodings that span more than one method.

In addition, the method could be evaluated on different kinds of bugs. There is also some room for improvement by balancing the training set with regards to the comparators in specific parts of code, like \textit{for loops}.
\section{Conclusion} \label{sec:conclusion}
We used encouraging results of Code2Vec model to test it on detecting off-by-one errors in arbitrary sized Java methods. The core idea of using a soft-attention mechanism to gain vector representations from AST paths remained the same. However, we modified the final layer to repurpose the model for detecting bugs instead of naming a method.

We tested different architectures of the model and tried to apply transfer learning. However, transfer learning proved only to be slightly better than a randomly initialized model and did not speed up the training process significantly. We trained the model on a large Java corpus of likely correct code to which we introduced simple mutations to get faulty samples.

The results of the quantitative and qualitative analyses show that the model has promising results and can generalize off-by-one errors in different contexts. However, the model suffers from the bias of the dataset and generates false positives for exotic code that deviates from standard style. In addition, the current model only analyses the AST of a single method, hence context is possibly lost which would allow detecting a bug.

We believe that this method could be tested with other bugs, hence all of our code and links to our training data are available at \url{https://github.com/serg-ml4se-2019/group5-deep-bugs}.

\bibliographystyle{plainnat}
\bibliography{main}

\newpage
\newgeometry{left=1cm,bottom=0cm,top=2cm,right=1cm}
\begin{appendices}
\twocolumn[\section{Qualitative Evaluation Code}\label{appendix:qualitative_evaluation}]

\begin{table}[htbp]
\begin{tabular}{|l|c|}
\hline
\textbf{Tool} & \textbf{Output} \\ \hline
Java Code & \begin{lstlisting}
public static void example() {
    int[] array = new int[5];
    for (int i = 0; i <= 5; i++) {
        System.out.println(array[i]);
    }
}
\end{lstlisting} \\ \hline
Gold & 1 (bug) \\ \hline
PVS-Studio & 1 \\ \hline
IntelliJ & 0 \\ \hline
SpotBugs & 0 \\ \hline
Our model & 1 \\ \hlineB{4}
Java Code & \begin{lstlisting}
public static void example() {
    int array[] = new int[6];
    for (int i = 0; i <= 5; i++) {
        System.out.println(array[i]);
    }
\end{lstlisting} \\ \hline
Gold & 0 (correct) \\ \hline
PVS-Studio & 0 \\ \hline
IntelliJ & 0 \\ \hline
SpotBugs & 0 \\ \hline
Our model & \textbf{1} (model wrong) \\ \hlineB{4}
Java Code & \begin{lstlisting}
public static void example() {
    int array[] = new int[5];
    for (int i = 0; i < 5; i++) {
        System.out.println(array[i]);
    }
}
\end{lstlisting} \\ \hline
Gold & 0 (correct) \\ \hline
PVS-Studio & 0 \\ \hline
IntelliJ & 0 \\ \hline
SpotBugs & 0 \\ \hline
Our model & 0 \\ \hlineB{4}
Java Code & \begin{lstlisting}
// Apache Tomcat Code
private boolean isPong() {
    return indexPong >= 0;
}
\end{lstlisting} \\ \hline
Gold & 0 (correct) \\ \hline
PVS-Studio & 0 \\ \hline
IntelliJ & 0 \\ \hline
SpotBugs & 0 \\ \hline
Our model & \textbf{1} (model wrong) \\ \hlineB{4}
Java Code & \begin{lstlisting}
// Apache Tomcat Code
public long getIdleTimeMillis() {
    final long elapsed = 
    System.currentTimeMillis() 
    - lastReturnTime;
    return elapsed >= 0 ? elapsed : 0;
}
\end{lstlisting} \\ \hline
Gold & 0 (correct) \\ \hline
PVS-Studio & 0 \\ \hline
IntelliJ & 0 \\ \hline
SpotBugs & 0 \\ \hline
Our model & \textbf{1} (style issue) \\ \hlineB{4}
\end{tabular}
\caption{Qualitative Evaluation Results}
\label{tab:manualEvaluationTable1}
\end{table}

\begin{table}[htbp]
\begin{tabular}{|l|c|}
\hline
\textbf{Tool} & \textbf{Output} \\ \hline
Java Code & \begin{lstlisting}
public boolean inBorders(int x, int y) {
    return x >= 0 && x <= getWidth() 
            && y >= 0 && y <= getHeight();
}
\end{lstlisting} \\ \hline
Gold & 0 (correct) \\ \hline
PVS-Studio & 0 \\ \hline
IntelliJ & 0 \\ \hline
SpotBugs & 0 \\ \hline
Our model & 0 \\ \hlineB{4}
Java Code & \begin{lstlisting}
public boolean inBorders(int x, int y) {
    return x > 0 && x <= getWidth() 
            && y >= 0 && y <= getHeight();
}
\end{lstlisting} \\ \hline
Gold & 1 (bug) \\ \hline
PVS-Studio & 0 \\ \hline
IntelliJ & 0 \\ \hline
SpotBugs & 0 \\ \hline
Our model & 1 \\ \hlineB{4}
Java Code & \begin{lstlisting}
public boolean contains(float value) {
    if (value > from && value <= to)
        return true;
    else
        return false;
}
\end{lstlisting} \\ \hline
Gold & 1 (bug) \\ \hline
PVS-Studio & 0 \\ \hline
IntelliJ & 0 \\ \hline
SpotBugs & 0 \\ \hline
Our model & 1 \\ \hlineB{4}
Java Code & \begin{lstlisting}
public boolean contains(float value) {
    if (value >= from && value <= to)
        return true;
    else
        return false;
}
\end{lstlisting} \\ \hline
Gold & 0 (correct) \\ \hline
PVS-Studio & 0 \\ \hline
IntelliJ & 0 \\ \hline
SpotBugs & 0 \\ \hline
Our model & 0 \\ \hlineB{4}
\end{tabular}
\caption{Qualitative Evaluation Results}
\label{tab:manualEvaluationTable2}
\end{table}

\onecolumn
\section{Quantitative Evaluation}\label{appendix:quantitative_evaluation}
\begin{table}[htbp]
\begin{tabular}{|l|r|r|r|r|r|r|r|r|r|}
\hline
\textbf{Context Type} & \multicolumn{1}{l|}{\textbf{TP}} & \multicolumn{1}{l|}{\textbf{TN}} & \multicolumn{1}{l|}{\textbf{FP}} & \multicolumn{1}{l|}{\textbf{FN}} & \multicolumn{1}{l|}{\textbf{Total}} & \multicolumn{1}{l|}{\textbf{Acc}} & \multicolumn{1}{l|}{\textbf{Recall}} & \multicolumn{1}{l|}{\textbf{Precision}} & \multicolumn{1}{l|}{\textbf{F1}} \\ \hline
FORlessEquals & \textbf{15906} & 177 & 573 & 2016 & \textbf{18672} & 0.8613 & \textbf{0.8875} & \textbf{0.9652} & \textbf{0.9247} \\ \hline
FORless & 214 & \textbf{16505} & 1417 & 536 & \textbf{18672} & \textbf{0.8954} & 0.2853 & 0.1312 & 0.1798 \\ \hline
IFgreaterEquals & 7835 & 3831 & \textbf{2253} & \textbf{3025} & 16944 & 0.6885 & 0.7215 & 0.7767 & 0.7480 \\ \hline
IFgreater & 3530 & 9290 & 1570 & 2554 & 16944 & 0.7566 & 0.5802 & 0.6922 & 0.6313 \\ \hline
IFlessEquals & 4498 & 900 & 735 & 1632 & 7765 & 0.6952 & 0.7338 & 0.8595 & 0.7917 \\ \hline
IFless & 605 & 5087 & 1043 & 1030 & 7765 & 0.7330 & 0.3700 & 0.3671 & 0.3686 \\ \hline
WHILEgreaterEquals & 813 & 237 & 150 & 251 & 1451 & 0.7236 & 0.7641 & 0.8442 & 0.8022 \\ \hline
WHILEgreater & 185 & 881 & 183 & 202 & 1451 & 0.7347 & 0.4780 & 0.5027 & 0.4901 \\ \hline
WHILElessEquals & 877 & 57 & 87 & 387 & 1408 & 0.6634 & 0.6938 & 0.9098 & 0.7873 \\ \hline
WHILEless & 33 & 1070 & 194 & 111 & 1408 & 0.7834 & 0.2292 & 0.1454 & 0.1779 \\ \hline
RETURNgreaterEquals & 553 & 284 & 104 & 245 & 1186 & 0.7057 & 0.6930 & 0.8417 & 0.7601 \\ \hline
RETURNgreater & 266 & 685 & 113 & 122 & 1186 & 0.8019 & 0.6856 & 0.7018 & 0.6936 \\ \hline
TERNARYgreaterEquals & 421 & 83 & 88 & 355 & 947 & 0.5322 & 0.5425 & 0.8271 & 0.6553 \\ \hline
TERNARYgreater & 77 & 627 & 149 & 94 & 947 & 0.7434 & 0.4503 & 0.3407 & 0.3879 \\ \hline
FORgreater & 488 & 129 & 122 & 98 & 837 & 0.7372 & 0.8328 & 0.8000 & 0.8161 \\ \hline
FORgreaterEquals & 107 & 515 & 71 & 144 & 837 & 0.7431 & 0.4263 & 0.6011 & 0.4988 \\ \hline
METHODgreaterEquals & 375 & 57 & 63 & 181 & 676 & 0.6391 & 0.6745 & 0.8562 & 0.7545 \\ \hline
METHODgreater & 42 & 441 & 115 & 78 & 676 & 0.7145 & 0.3500 & 0.2675 & 0.3032 \\ \hline
TERNARYlessEquals & 304 & 49 & 56 & 201 & 610 & 0.5787 & 0.6020 & 0.8444 & 0.7029 \\ \hline
TERNARYless & 48 & 437 & 68 & 57 & 610 & 0.7951 & 0.4571 & 0.4138 & 0.4344 \\ \hline
RETURNlessEquals & 302 & 120 & 79 & 92 & 593 & 0.7116 & 0.7665 & 0.7927 & 0.7794 \\ \hline
RETURNless & 111 & 338 & 56 & 88 & 593 & 0.7572 & 0.5578 & 0.6647 & 0.6066 \\ \hline
METHODlessEquals & 111 & 83 & 50 & 57 & 301 & 0.6445 & 0.6607 & 0.6894 & 0.6748 \\ \hline
METHODless & 42 & 129 & 39 & 91 & 301 & 0.5681 & 0.3158 & 0.5185 & 0.3925 \\ \hline
ASSERTgreaterEquals & 56 & 66 & 16 & 48 & 186 & 0.6559 & 0.5385 & 0.7778 & 0.6364 \\ \hline
ASSERTgreater & 36 & 70 & 34 & 46 & 186 & 0.5699 & 0.4390 & 0.5143 & 0.4737 \\ \hline
VARIABLEDECLARATORgreaterEquals & 104 & 26 & 7 & 47 & 184 & 0.7065 & 0.6887 & 0.9369 & 0.7939 \\ \hline
VARIABLEDECLARATORgreater & 7 & 124 & 27 & 26 & 184 & 0.7120 & 0.2121 & 0.2059 & 0.2090 \\ \hline
DOgreaterEquals & 86 & 13 & 10 & 44 & 153 & 0.6471 & 0.6615 & 0.8958 & 0.7611 \\ \hline
DOgreater & 8 & 111 & 19 & 15 & 153 & 0.7778 & 0.3478 & 0.2963 & 0.3200 \\ \hline
DOlessEquals & 78 & 2 & 4 & 62 & 146 & 0.5479 & 0.5571 & 0.9512 & 0.7027 \\ \hline
DOless & 4 & 110 & 30 & 2 & 146 & 0.7808 & 0.6667 & 0.1176 & 0.2000 \\ \hline
ASSIGNgreaterEquals & 51 & 18 & 12 & 59 & 140 & 0.4929 & 0.4636 & 0.8095 & 0.5896 \\ \hline
ASSIGNgreater & 11 & 87 & 23 & 19 & 140 & 0.7000 & 0.3667 & 0.3235 & 0.3438 \\ \hline
ASSERTlessEquals & 56 & 23 & 23 & 16 & 118 & 0.6695 & 0.7778 & 0.7089 & 0.7417 \\ \hline
ASSERTless & 13 & 57 & 15 & 33 & 118 & 0.5932 & 0.2826 & 0.4643 & 0.3514 \\ \hline
VARIABLEDECLARATORlessEquals & 25 & 20 & 11 & 32 & 88 & 0.5114 & 0.4386 & 0.6944 & 0.5376 \\ \hline
VARIABLEDECLARATORless & 11 & 42 & 15 & 20 & 88 & 0.6023 & 0.3548 & 0.4231 & 0.3860 \\ \hline
ASSIGNlessEquals & 9 & 4 & 2 & 13 & 28 & 0.4643 & 0.4091 & 0.8182 & 0.5455 \\ \hline
ASSIGNless & 0 & 17 & 5 & 6 & 28 & 0.6071 & 0.0000 & 0.0000 & 0.0000 \\ \hline
EXPRESSIONgreaterEquals & 10 & 2 & 1 & 13 & 26 & 0.4615 & 0.4348 & 0.9091 & 0.5882 \\ \hline
EXPRESSIONgreater & 0 & 22 & 1 & 3 & 26 & 0.8462 & 0.0000 & 0.0000 & 0.0000 \\ \hline
EXPRESSIONlessEquals & 5 & 1 & 2 & 2 & 10 & 0.6000 & 0.7143 & 0.7143 & 0.7143 \\ \hline
EXPRESSIONless & 0 & 4 & 3 & 3 & 10 & 0.4000 & 0.0000 & 0.0000 & 0.0000 \\ \hline
OBJECTCREATIONgreaterEquals & 3 & 1 & 0 & 4 & 8 & 0.5000 & 0.4286 & \textbf{1.0000} & 0.6000 \\ \hline
OBJECTCREATIONgreater & 0 & 5 & 2 & 1 & 8 & 0.6250 & 0.0000 & 0.0000 & 0.0000 \\ \hline
OBJECTCREATIONlessEquals & 1 & 0 & 0 & 1 & 2 & 0.5000 & 0.5000 & \textbf{1.0000} & 0.6667 \\ \hline
OBJECTCREATIONless & 0 & 1 & 1 & 0 & 2 & 0.5000 & 0.0000 & 0.0000 & 0.0000 \\ \hline
 &  &  &  &  &  &  &  &  &  \\ \hline
Total & 38317 & 42838 & 9641 & 14162 & 104958 & 0.7732 & 0.7301 & 0.7990 & 0.7630 \\ \hline
\end{tabular}
\caption{Quantitative evaluation results for all context types analyzed}
\label{tab:quantitative_evaluation_all_bugs}
\end{table}

\begin{table*}[htbp]
\begin{tabular}{|l|r|r|r|r|r|r|r|r|r|}
\hline
\textbf{Statement Type for Bug} & \multicolumn{1}{l|}{\textbf{TP}} & \multicolumn{1}{l|}{\textbf{TN}} & \multicolumn{1}{l|}{\textbf{FP}} & \multicolumn{1}{l|}{\textbf{FN}} & \multicolumn{1}{l|}{\textbf{Total}} & \multicolumn{1}{l|}{\textbf{Acc}} & \multicolumn{1}{l|}{\textbf{Recall}} & \multicolumn{1}{l|}{\textbf{Precision}} & \multicolumn{1}{l|}{\textbf{F1}} \\ \hline
IF & 16471 & \textbf{19108} & \textbf{5601} & \textbf{8238} & \textbf{49418} & 0.72 & 0.6666 & 0.7462 & 0.7042 \\ \hline
FOR & \textbf{16709} & 17326 & 2183 & 2800 & 39018 & \textbf{0.8723} & \textbf{0.8565} & \textbf{0.8844} & \textbf{0.8702} \\ \hline
WHILE & 1913 & 2245 & 614 & 946 & 5718 & 0.7272 & 0.6691 & 0.757 & 0.7104 \\ \hline
RETURN & 1233 & 1427 & 352 & 546 & 3558 & 0.7476 & 0.6931 & 0.7779 & 0.7331 \\ \hline
TERNARY & 851 & 1196 & 361 & 706 & 3114 & 0.6574 & 0.5466 & 0.7021 & 0.6147 \\ \hline
METHOD & 578 & 710 & 267 & 399 & 1954 & 0.6592 & 0.5916 & 0.684 & 0.6345 \\ \hline
ASSERT & 157 & 216 & 88 & 147 & 608 & 0.6135 & 0.5164 & 0.6408 & 0.5719 \\ \hline
DO & 176 & 236 & 63 & 123 & 598 & 0.689 & 0.5886 & 0.7364 & 0.6543 \\ \hline
VARIABLEDECLARATOR & 146 & 212 & 60 & 126 & 544 & 0.6581 & 0.5368 & 0.7087 & 0.6109 \\ \hline
ASSIGN & 75 & 126 & 42 & 93 & 336 & 0.5982 & 0.4464 & 0.641 & 0.5263 \\ \hline
EXPRESSION & 15 & 29 & 7 & 21 & 72 & 0.6111 & 0.4167 & 0.6818 & 0.5172 \\ \hline
OBJECTCREATION & 4 & 7 & 3 & 6 & 20 & 0.55 & 0.4 & 0.5714 & 0.4706 \\ \hline
 & \multicolumn{1}{l|}{} & \multicolumn{1}{l|}{} & \multicolumn{1}{l|}{} & \multicolumn{1}{l|}{} & \multicolumn{1}{l|}{} & \multicolumn{1}{l|}{} & \multicolumn{1}{l|}{} & \multicolumn{1}{l|}{} & \multicolumn{1}{l|}{} \\ \hline
Total  & 38328 & 42838 & 9641 & 14151 & 104958 & 0.7733 & 0.7303 & 0.799 & 0.7631 \\ \hline
\end{tabular}
\caption{Quantitative analysis by statement type}
\label{tab:quantitative_evaluation_bug_types}
\end{table*}

\begin{table*}[htbp]
\begin{tabular}{|l|r|r|r|r|r|r|r|r|r|}
\hline
\textbf{Comparator} & \multicolumn{1}{l|}{\textbf{TP}} & \multicolumn{1}{l|}{\textbf{TN}} & \multicolumn{1}{l|}{\textbf{FP}} & \multicolumn{1}{l|}{\textbf{FN}} & \multicolumn{1}{l|}{\textbf{Total}} & \multicolumn{1}{l|}{\textbf{Accuracy}} & \multicolumn{1}{l|}{\textbf{Recall}} & \multicolumn{1}{l|}{\textbf{Precision}} & \multicolumn{1}{l|}{\textbf{F1}} \\ \hline
LessEquals & \textbf{22172} & 1436 & 1622 & 4511 & \textbf{29741} & 0.7938 & \textbf{0.8309} & \textbf{0.9318} & \textbf{0.8785} \\ \hline
Less & 1081 & \textbf{23797} & \textbf{2886} & 1977 & \textbf{29741} & \textbf{0.8365} & 0.3535 & 0.2725 & 0.3078 \\ \hline
Greater & 4650 & 12472 & 2358 & 3258 & 22738 & 0.753 & 0.588 & 0.6635 & 0.6235 \\ \hline
GreaterEquals & 10414 & 5133 & 2775 & \textbf{4416} & 22738 & 0.6522 & 0.7022 & 0.7896 & 0.7434 \\ \hline
 & \multicolumn{1}{l|}{} & \multicolumn{1}{l|}{} & \multicolumn{1}{l|}{} & \multicolumn{1}{l|}{} & \multicolumn{1}{l|}{} & \multicolumn{1}{l|}{} & \multicolumn{1}{l|}{} & \multicolumn{1}{l|}{} & \multicolumn{1}{l|}{} \\ \hline
Total & 38317 & 42838 & 9641 & 14162 & 104958 & 0.7732 & 0.7301 & 0.799 & 0.763 \\ \hline
\end{tabular}
\caption{Quantitative analysis by comparator type}
\label{tab:operator_statistics}
\end{table*}

\begin{table*}[htbp]
\begin{tabular}{|l|r|r|r|r|r|r|r|r|r|}
\hline
\textbf{Context Type} & \multicolumn{1}{l|}{\textbf{TP}} & \multicolumn{1}{l|}{\textbf{TN}} & \multicolumn{1}{l|}{\textbf{FP}} & \multicolumn{1}{l|}{\textbf{FN}} & \multicolumn{1}{l|}{\textbf{Total}} & \multicolumn{1}{l|}{\textbf{Accuracy}} & \multicolumn{1}{l|}{\textbf{Recall}} & \multicolumn{1}{l|}{\textbf{Precision}} & \multicolumn{1}{l|}{\textbf{F1}} \\ \hline
IFgreaterEquals & \textbf{7978} & 3647 & \textbf{2460} & 2845 & \textbf{16930} & \textbf{0.6867} & 0.7371 & 0.7643 & 0.7505 \\ \hline
IFgreater & 1744 & \textbf{8428} & 2395 & \textbf{4363} & \textbf{16930} & 0.6008 & 0.2856 & 0.4214 & 0.3404 \\ \hline
IFlessEquals & 4447 & 736 & 908 & 1688 & 7779 & 0.6663 & 0.7249 & \textbf{0.8304} & \textbf{0.7741} \\ \hline
IFless & 539 & 4825 & 1310 & 1105 & 7779 & 0.6895 & 0.3279 & 0.2915 & 0.3086 \\ \hline
 & \multicolumn{1}{l|}{} & \multicolumn{1}{l|}{} & \multicolumn{1}{l|}{} & \multicolumn{1}{l|}{} & \multicolumn{1}{l|}{} & \multicolumn{1}{l|}{} & \multicolumn{1}{l|}{} & \multicolumn{1}{l|}{} & \multicolumn{1}{l|}{} \\ \hline
Total & 14708 & 17636 & 7073 & 10001 & 49418 & 0.6545 & 0.5952 & 0.6753 & 0.6327 \\ \hline
\end{tabular}
\caption{Quantitative evaluation results for a model trained only on bugs concerned with \textit{if} statements}
\label{tab:quantitative_if_evaluation}
\end{table*}

\end{appendices}

\end{document}